\documentclass[aps,pre,twocolumn,superscriptaddress,groupedaddress,showpacs]{revtex4}
\usepackage{amsmath,amssymb}
\usepackage[dvipdfm]{graphicx}
\begin{document}
%\preprint{}
\title{Loewner driving functions for off-critical  percolation clusters}
\author{Yoichiro Kondo}
\email[]{ykondo@stat.phys.kyushu-u.ac.jp}
\affiliation{Department of Physics, Kyushu University, 33, Fukuoka 812-8581, Japan}
\author{Namiko Mitarai}
\affiliation{Niels Bohr Institute, Blegdamsvej 17, DK-2100, Copenhagen, Denmark}
\author{Hiizu Nakanishi}
\affiliation{Department of Physics, Kyushu University, 33, Fukuoka 812-8581, Japan}

%\homepage[]{http://www.stat.phys.kyushu-u.ac.jp/~ykondo/}
%\thanks{}

\date{\today}
\begin{abstract}
We numerically study the Loewner driving function $U_t$ of a site
percolation cluster boundary on the triangular lattice for $p<p_c$.  It
is found that $U_t$ shows a drifted random walk with a finite
crossover time.  Within this crossover time, the averaged driving
function $\left< U_t\right>$ shows a scaling behavior $-(p_c-p)\,
t^{(\nu +1)/2\nu}$ with a superdiffusive fluctuation whereas, beyond the
crossover time, the driving function $U_t$ undergoes a normal diffusion 
with Hurst exponent $1/2$ but with the drift velocity proportional to $(p_c-p)^\nu$, where
$\nu= 4/3$ is the critical exponent for two-dimensional percolation
correlation length.  The crossover time diverges as $(p_c-p)^{-2\nu}$ as
$p\to p_c$.
\end{abstract}
%\keywords{}
\maketitle
\textit{Introduction.}
Loewner evolution has recently drawn much attention in physics community
because of the development of  
Schramm-Loewner evolution (SLE) \cite{Schramm99}, which has provided us
a new tool for the study of two-dimensional (2-d) continuous phase transition.
The basic device of SLE is a conformal mapping that transforms a motion
along a non-intersecting 2-d curve into another motion along
the real axis; \textit{Loewner driving function} is a real function that
represents this transformed motion. It turns out that, for a certain class of stochastic and conformally
invariant curves in 2-d, the driving
function shows Brownian motion in one dimension.
What makes SLE especially remarkable is that it gives us a method that describes
all the geometrical properties of the curves through a single parameter of the Brownian
motion, namely, the diffusion constant. 
The class of curves  includes the self-avoiding walk, the uniform-spanning trees, the loop-erased random walk, and boundaries of critical clusters in various
 2-d lattice models in physics such as percolation, Ising
 model, $O(n)$ loop models, and Potts models (see
 \cite{Lawler} for review).

Being inspired by mathematically oriented development, people
start using SLE formalism to test the conformal invariance by
calculating Loewner driving functions obtained
from 2-d curves in a number of physical systems, such as
vorticity clusters and temperature isolines in turbulence \cite{Turb1},
domain walls in 2-d spin glass \cite{SG1}, isoheight
lines on growing solid
surface \cite{Growth1}, nodal domains of chaotic maps \cite{Chaos1}, etc.

In study of physical systems, it is important to ask how driving
functions may look like when the system
departs from the critical point, because in real life there are a number of sources
that may drive a system away from it. There are some mathematical approaches to study
the effects of off-criticality on SLE based upon probability theory and conformal
field theory \cite{Nolin,off-sles}, which mainly pursue mathematical
consistency in the continuum limit, but general feature of the
off-critical driving function has not been known yet.
In this Letter, we report our results of numerical simulations to
study how the Loewner driving function deviates from the ideal Brownian
motion when the system departs from the criticality in the case of percolation clusters.

\begin{figure}[thb]
 \includegraphics[width=75mm]{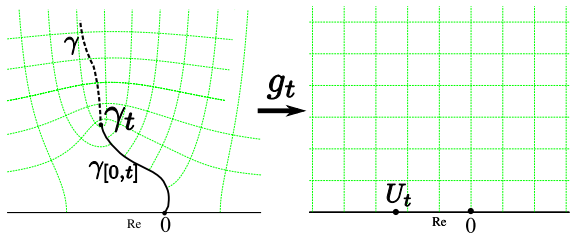}
 \caption{(Color online) A schematic diagram of conformal mapping. $g_t$ maps
 $\mathbb{H}\setminus \gamma_{(0,t]}$ to $\mathbb{H}$ with $\gamma_t$ to
 $U_t$ on the real axis. The deformed grid in the left
 panel can be regarded as equipotential and electric force lines, and
 is mapped to the straight grid in the right panel. \label{fig:gamma-sample}}
\end{figure}

\textit{Loewner evolution.}
 Let us start by reviewing basic elements of Loewner evolution briefly. Consider a
 non-intersecting continuous curve $\gamma$ which starts from the origin
and extends toward infinity in the upper half plane $\mathbb{H}$. We
 parametrize $\gamma$ by $t \ge 0$, and denote a point on $\gamma$ as
 $\gamma_t$ with $\gamma_0 = 0$. A part of $\gamma$ between
 $\gamma_{t_1}$ and $\gamma_{t_2}$ is represented by $\gamma_{[t_1, t_2]}$.
It is known that, for given $\gamma_{[0,t]}$, there exists a unique
 conformal map $g_t: \mathbb{H} \setminus \gamma_{(0,t]} \to \mathbb{H}$ that
 satisfies the condition
\begin{eqnarray}
  g_t(z) = z + a_t/z + O(|z|^{-2}) \hspace{1em}\: \mbox{as} \:  |z| \to \infty
\end{eqnarray}
with $a_t \ge 0$ (Fig.\ \ref{fig:gamma-sample}). The parameter $t$, which
we will call time, is now defined by $t := a_t/2$. The driving function
$U_t$ is defined by the image of $\gamma_t$ by the map $g_t$:
%
%\begin{eqnarray}
 $ U_t := \lim_{z\to\gamma_t} g_t(z). $
%\end{eqnarray} 
%
Then, it can be shown that $g_t(z)$ satisfies the Loewner evolution,
%\begin{eqnarray}
 $\partial_t{g_t(z)} = 2/(g_t(z) - U_t).\label{eq:le}$
%\end{eqnarray}

Note that $g_t(z)$ represents the complex electrostatic
potential for the equipotential boundary of $\gamma_{[0,t]}$ and the
real axis \cite{Landau}, then one can see that the time $a_t = 2t$ is equal to
$p/2\pi$, where $p$ is the 2-d dipole moment
induced by $\gamma_{[0,t]}$ on a flat electrode. In this electro-static
picture, $U_t$ is just given by the charge induced by $\gamma_{[0,t]}$
along the right side of $\gamma_{[0,t]}$ and the positive part of the
real axis in the unit system with $\varepsilon_0 = 1$ (Fig.\ \ref{fig:gamma-sample}).

Schramm has shown that $U_t$ becomes a Brownian motion if the curve $\gamma$
is a conformally invariant random curve with the domain Markov
 property \cite{Schramm99}. This means that any
 properties of the curve such as the fractal dimension
 are determined solely by the diffusion constant of the Browninan motion.

\begin{figure}[bt]
\includegraphics[width=75mm]{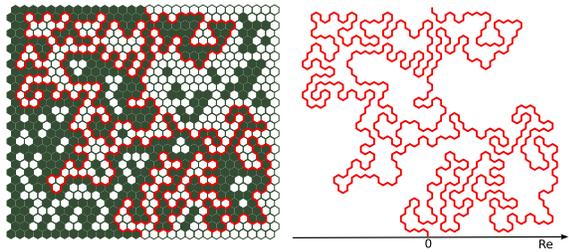}
 \caption{(Color online) Percolation on the triangular lattice. The left panel shows a
 whole system with left (right) half of the system boundary being occupied (unoccupied). The right panel shows only the cluster boundary that
 starts from the origin.\label{fig:exp}}
\end{figure}

\textit{Simulations.}
 We performed numerical simulations to obtain $U_t$ for cluster boundaries of the site
 percolation on the triangular lattice with the occupation probability
 $p \le p_c =0.5$. Percolation clusters are generated in the
 rectangular system. To ensure the
 cluster boundaries to extend from the origin in the upper half plane, the
 left (right) half of the boundary sites are set to be occupied (unoccupied),
 and the boundaries are defined along the edges of the dual lattice,
 i.e., the honeycomb
 lattice \cite{Lawler} (Fig.\ \ref{fig:exp}). We use only the
 part of cluster boundaries that never touches the peripheries except for the
 bottom side. %The correlation length $\xi$ is estimated by taking twice the average
%height of $\gamma$ where $p_c - p > 0.04$  and extrapolating data toward $p_c$.
For each boundary curve, we numerically generate the conformal
 map $g_t(z)$ and compute the driving function $U_t$ using the zipper
 algorithm \cite{Kennedy-zipper,Marshall}. 
\begin{figure}[bth]
\begin{center}
\includegraphics[width=75mm]{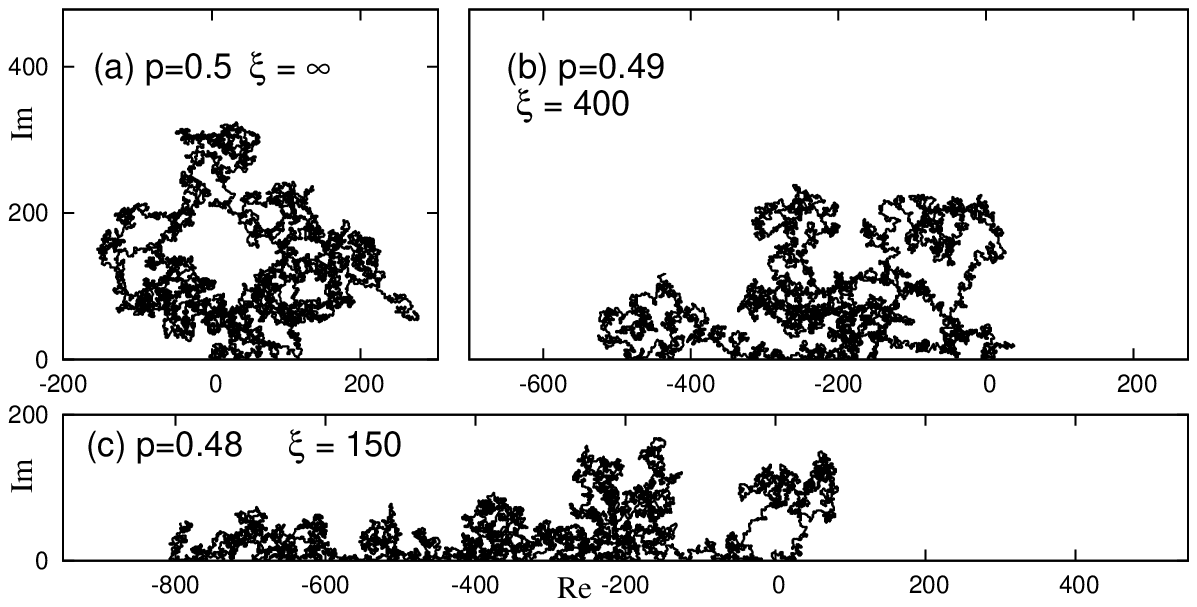}\\
\includegraphics[width=75mm]{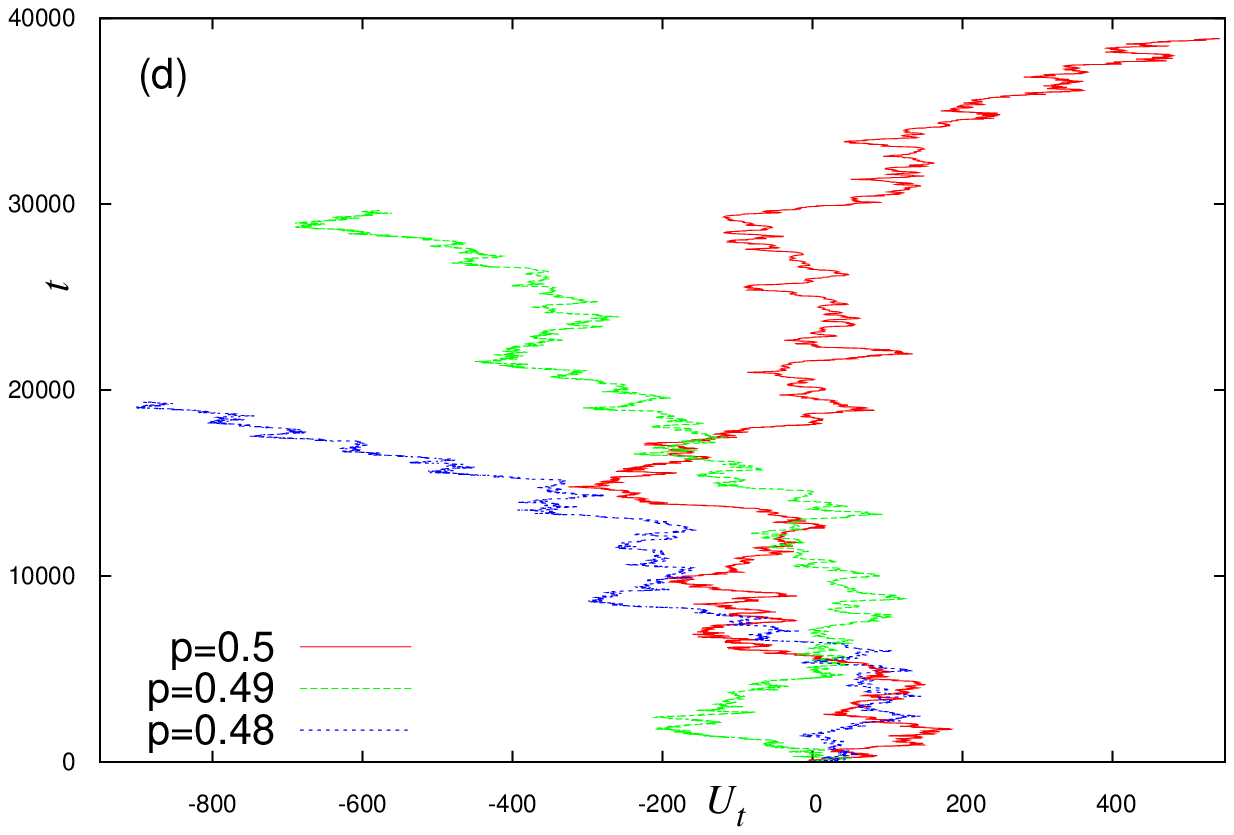}
 \caption{(Color online) Examples of cluster boundaries (a-c) and their driving
 functions (d) for $p=0.5, 0.49$, and $0.48$. The boundaries are shown up
 to $N=20000$ steps. The lattice spacing is set to be unity. The correlation lengths are $\xi = 400$ and $150$ for
 $p=0.49$ and $0.48$, respectively. 
 \label{fig:ut-sample}}
\end{center}
\end{figure}

 Fig.\ \ref{fig:ut-sample}(a-c) shows typical examples of boundaries
 $\gamma$ up to $20000$ steps at $p=0.5, 0.49,$ and $0.48$, and
 Fig.\ \ref{fig:ut-sample}(d) shows corresponding driving functions
 $U_t$ \cite{note}.
 For $p=0.5$, the boundary extends into $\mathbb{H}$ indefinitely
 around the origin once it hits an infinite cluster. For $p<p_c$, the boundary tends to extend toward
 left in larger scale, but it can hardly be distinguished from that at
 $p=p_c$ within the scale of correlation length $\xi$ with a finite
 lattice constant.
 Accordingly, $U_t$ wander around the origin at $p = p_c$ whereas it drifts toward
 $-\infty$ for large $t$ for $p<p_c$. Note that the time does not increases
 uniformly with the step along a boundary because the time is proportional to the
 dipole moment induced by $\gamma_{[0,t]}$.

\begin{figure}[t]
\begin{center}
 \includegraphics[width=90mm]{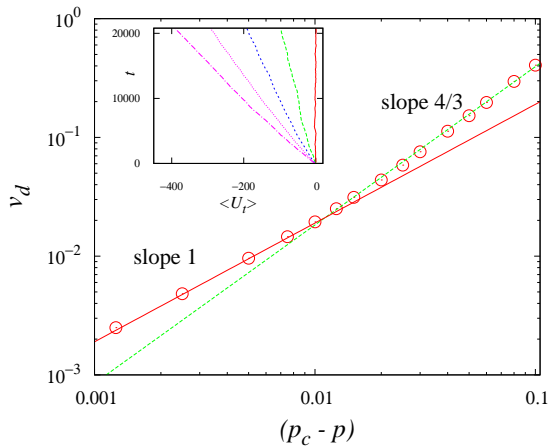}
 \caption{(Color online) Drift velocity $v_d$ vs. $(p_c -p)$. The drift velocities of
 the driving function averaged up to $t=21000$ are plotted in the logarithmic scale as a function
 of $(p_c - p)$. There is a crossover around $(p_c - p) \sim 0.01$ between the
 two regimes. The solid (dashed) line with the slope $1$ ($4/3$) shows
 the limiting behavior in the small (large) $p_c-p$ regime. The
 inset shows $t$ vs. $\langle U_t \rangle$ for $p = 0.5, 0.4975, 0.495, 0.4925$ and $0.49$ (from
 right to left). Each line represents average behavior over $4000$ samples.\label{fig:drift}}
\end{center}
\end{figure}

The averaged behavior of $U_t$ is shown for several values of $p$ in the
inset of Fig.\ \ref{fig:drift}, where one can see $\langle{U_t}\rangle$
appears to drift at a constant velocity $v_d$. In Fig.\ \ref{fig:drift},
the drift velocity $v_d$ averaged up to $t=21000$ over $4000$ samples is
plotted against $p_c - p$ in the logarithmic scale;
$v_d$ is shown to be proportional to $(p_c - p)$ for $(p_c - p) \lesssim
 0.01$,
\begin{eqnarray}
 v_d \sim (p_c - p),
\end{eqnarray} 
whereas it behaves as
\begin{eqnarray}
 v_d \sim (p_c - p)^{\nu}
\end{eqnarray} 
with $\nu \approx 4/3 $ for $p_c - p \gtrsim 0.01$.

\begin{figure}[t]
\begin{center}
\includegraphics[width=90mm]{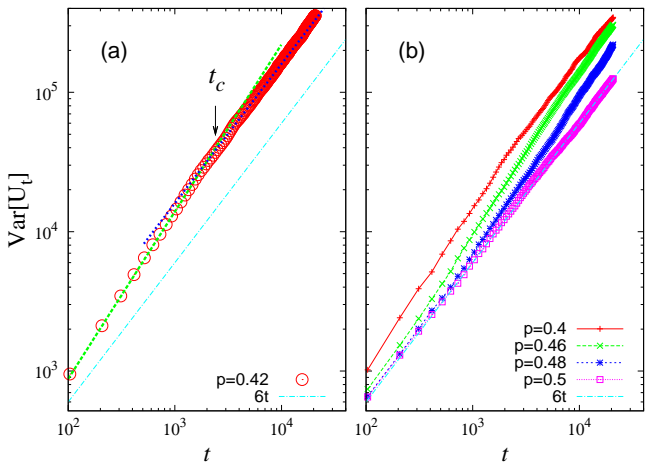}
\caption{(Color online) Var[$U_t$] vs. $t$ in the logarithmic scale. (a) The plot for
 $p=0.42$. The behavior changes from superdiffusive to diffusive around
 the crossover time $t_c$ indicated by the arrow. The behavior 
 for $p=p_c$,
 Var[$U_t$]=$6t$, is plotted for comparison.  (b) The plots for several
 values of $p$. The crossover time
$t_c$ becomes larger as we approach $p_c$. \label{fig:var}}
\end{center}
\end{figure}
\begin{figure}[tb]
\begin{center}
\includegraphics[width=90mm]{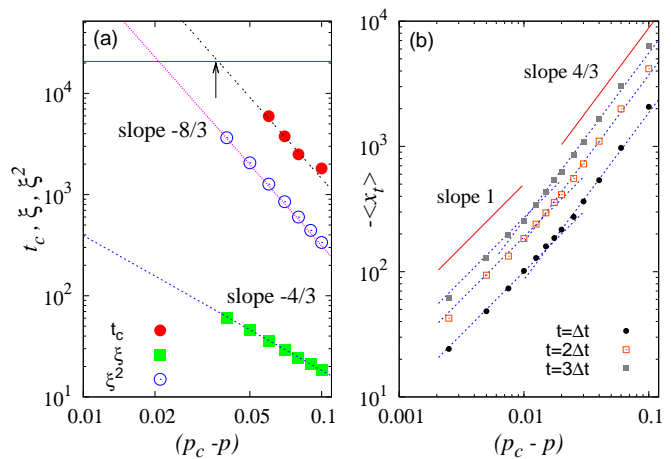}
\caption{(Color online) (a)$t_c$, $\xi$, and $\xi^2$  vs. $(p_c - p)$ in
 the logarithmic scale. The horizontal solid line at $t=21000$ indicates
 the simulation range. 
(b) $-\langle x_t \rangle$ vs. $(p_c -p)$ for  $t=\Delta{t}, 2\Delta{t},$
 and $3\Delta{t}$ with $\Delta{t} = 5200$, in the logarithmic scale. Each data point represents  average over $3000$ samples. \label{fig:xi2ave}}
\end{center}
\end{figure}

The time dependence of the variance Var[$U_t$] is shown in the
logarithmic scale in Fig.\ \ref{fig:var}. As one can see in the plot for
$p=0.42$ in Fig.\ \ref{fig:var}(a), there are two time regimes: the superdiffusion regime
and  the normal diffusion regime,
\begin{eqnarray}
 \mathrm{Var}[U_t] \sim \left\{ 
			 \begin{array}{ll}
			 t^\alpha & \mbox{for } t\ll t_c\\
			  t & \mbox{for } t\gg t_c
\end{array}
\right. ,
\end{eqnarray}
with the exponent $\alpha > 1$ and the crossover time $t_c$. The plots
for several values of $p$ in Fig.\ \ref{fig:var}(b) shows the tendency that the exponent $\alpha$
decreases to $1$ and the crossover time $t_c$ increases as $p \to
p_c$. The estimated values for $t_c$ are plotted against $(p_c - p)$ in
the logarithmic scale
by filled circles in Fig.\ \ref{fig:xi2ave}(a).

\textit{Discussions.} 
Our results can be understood as in the following.  First, we consider
the curve of $p = p_c$. In this case, $\gamma_t$ explores along a critical
boundary, which extends indefinitely without bias. Consider a part of
the curve $\gamma_{[0,t]}$. This may look like a blob, whose size we
denote by $l_t$.
The time $t$ that corresponds to the blob can be estimated as
\begin{eqnarray}
 t \sim l_t^2,
\end{eqnarray}
using the electro-static analogy; The time $t$ is proportional to the dipole
moment $p$ induced by $\gamma_{[0,t]}$ attached to a flat electrode, and
both the induced charge and the charge displacement are of order of
$l_t$. On the other hand, $U_t$ may be approximated to be $x_t:=\mbox{Re}\gamma_t$ because
$U_t$ is determined from the number of electric force lines landing on
the right side of $\gamma_{[0,t]}$ and the positive part of real axis. Since $x_t$ wanders along the blob of size
$l_t$, one can see that $U_t$ undergoes normal diffusion without drift:
\begin{eqnarray}
 U_t \sim l_t \sim \sqrt{t}.
\end{eqnarray}

For $p < p_c$, the curve
$\gamma$ is the boundary of finite off-critical clusters that are connected by
the bottom. Since typical size of each cluster is the
percolation correlation length $\xi$, the curve $\gamma$
looks like a chain of the blobs of size $\xi$ (see for example Fig.\ $\ref{fig:ut-sample}$(c)). Within each blob, $\gamma$ is almost like a critical boundary
but with a bias toward the left; the strength of the bias is proportional to $p_c - p$.

There should be two time regimes: the short time regime
where $\gamma_t$ is still in the first blob, and the long time regime
where $\gamma_t$ is traveling over blobs.  The crossover time $t_c$
between the two regimes is the time that $\gamma_t$ goes over the first
blob of size $\xi$, therefore, it is estimated as $t_c \sim \xi^2$ as in
the case of critical blob.  Within the short time regime, $\gamma_t$
explores in the first blob that looks almost like a critical blob in the
smaller length scale than $\xi$ but with a small bias toward left.  We
assume the scaling behavior $\left< x_t \right>\sim -\xi\,(t/t_c)^\beta$
with an exponent $\beta$. If we determine the exponent so that the
effect of bias should be proportional to $(p_c - p)$ for a fixed $t$,
we obtain $\beta=(\nu+1)/2\nu=7/8$, using the critical exponent
$\nu=4/3$ for the correlation length of the 2-d percolation.
On the other hand,
the behavior of $\left< x_t\right>$ in the long time regime may be
obtained by considering the situation where $\gamma_t$ is at $n$-th
blob, i.e., $\left< x_t\right>\sim n\xi$.  The time corresponding to
this is $t\sim n\xi^2$ because the dipole moment induced at each blob is
of order of $\xi^2$, with which we can express $\left< x_t\right>$ as
$t/\xi$.
With all these argument, we finally obtain
\begin{equation}
\left< x_t \right>  \sim
\left\{\begin{array}{ll}
-(p_c-p)\, t^{(\nu+1)/2\nu}  & (t\lesssim t_c)
\\\\
-(p_c-p)^{\nu}\, t & (t\gtrsim t_c)
\end{array}
\right.
\label{x_t}
\end{equation}
with
%\begin{equation}
$t_c\sim (p_c-p)^{-2\nu}.
\label{t_c}$
%\end{equation}

Within the approximation $U_t\sim x_t$, this is consistent with our
results for the drift velocity $v_d$ in Fig.\ \ref{fig:drift}; The slope we obtained for
$p_c-p\gtrsim 0.01$ is very close to $\nu=4/3$.  The linear dependence for
$p_c-p\lesssim 0.01$ should correspond to the short time behavior. The
crossover around $p_c - p \approx 0.01$ is due to the fact that our
simulation time  length is not long enough  in comparison with $t_c$. If
we can simulate longer time, the crossover value of $p_c - p$ will be smaller.
The expected weak non-linear
$t$-dependence in $x_t$ is difficult to
distinguish from the linear behavior numerically; Actually, one might
notice slight convexity of the plots in the inset of Fig.\ \ref{fig:drift}.

In order to check further consistency of the data with our
interpretation, we plot $t_c$ for ${\rm Var}[U_t]$, $\xi$, and $\xi^2$ against $p_c-p$ in Fig.\ \ref{fig:xi2ave}(a).  The horizontal solid line indicates
the time range of our simulations.
The plotted range of the crossover time (filled circles) in ${\rm
Var}[U_t]$ is too narrow to determine its behavior in a reliable way,
but it seems consistent with that of $\xi^2$ (open circles).  The value
of $p_c-p$ where the extrapolated $t_c$ of ${\rm Var}[U_t]$ reaches the
simulation time is about 0.04 (arrow), which is larger than the
crossover point in $v_d$ by the factor 4; This discrepancy may come from
the difference between $v_d$ and Var[$U_t$], or simply due to the
uncertainty in the estimates.
In Fig.\ \ref{fig:xi2ave}(b), $-\left< x_t\right>$ for several values of $t$ are plotted
against $(p_c-p)$.  From Eq.\ (\ref{x_t}), we expect linear dependence on
$(p_c-p)$ for $(p_c-p)\lesssim t^{-3/8}$ and $(p_c-p)^{4/3}$-dependence
beyond that; 
The data seem to be consistent.

With these observations, we interpret our results as follows. For an
off-critical percolation cluster boundary, the driving function $U_t$
undergoes a random walk with drift. The drift velocity $v_d$ appears to
be proportional to $(p_c -p)$ when $(p_c - p) \lesssim 0.01$, but it is
because time lengths of our simulations are finite; With the approximation $U_t \sim x_t$, the averaged driving function
 should be given by Eq.\ (\ref{x_t}). Within the crossover time,
the random walk is superdiffusive with exponent $\alpha > 1$, which decreases toward $1$ as $p \to p_c$. Beyond the crossover time, the
fluctuation around the drift motion is normal diffusion with larger
diffusion constant than that at $p=p_c$.

Finally, let us discuss the scaling limit where the lattice constant
$a\to 0$ with keeping the correlation length $\xi$ constant.  Our
argument to derive Eq.(7) holds for any $\xi$ and $a$ as long as
$\xi\gg a$, thus if we scale the variables as
$
\tilde x_t \equiv x_t/\xi,\,
\tilde U_t \equiv U_t/\xi,\,$and $
\tilde t \equiv t/\xi^2,
$
then we obtain the corresponding equation for the scaled variables, i.e.
the same as Eq.(7) but without the factor of $(p_c-p)$.


\begin{thebibliography}{20}
\expandafter\ifx\csname natexlab\endcsname\relax\def\natexlab#1{#1}\fi
\expandafter\ifx\csname bibnamefont\endcsname\relax
  \def\bibnamefont#1{#1}\fi
\expandafter\ifx\csname bibfnamefont\endcsname\relax
  \def\bibfnamefont#1{#1}\fi
\expandafter\ifx\csname citenamefont\endcsname\relax
  \def\citenamefont#1{#1}\fi
\expandafter\ifx\csname url\endcsname\relax
  \def\url#1{\texttt{#1}}\fi
\expandafter\ifx\csname urlprefix\endcsname\relax\def\urlprefix{URL }\fi
\providecommand{\bibinfo}[2]{#2}
\providecommand{\eprint}[2][]{\url{#2}}

\bibitem[{\citenamefont{Schramm}(2000)}]{Schramm99}
\bibinfo{author}{\bibfnamefont{O.}~\bibnamefont{Schramm}},
  \bibinfo{journal}{Isr. J. Math.} \textbf{\bibinfo{volume}{118}},
  \bibinfo{pages}{221} (\bibinfo{year}{2000}).

\bibitem[{\citenamefont{Lawler}(2005)}]{Lawler}
\bibinfo{author}{\bibfnamefont{G.}~\bibnamefont{Lawler}},
  \emph{\bibinfo{title}{Conformally Invariant Processes in the Plane}}
  (\bibinfo{publisher}{American Mathematical Society, Providence},
  \bibinfo{year}{2005});
%\bibitem[{\citenamefont{Kager and Nienhuis}(2004)}]{Kager-review}
\bibinfo{author}{\bibfnamefont{W.}~\bibnamefont{Kager}} \bibnamefont{and}
  \bibinfo{author}{\bibfnamefont{B.}~\bibnamefont{Nienhuis}},
  \bibinfo{journal}{J. Stat. Phys.} \textbf{\bibinfo{volume}{115}},
  \bibinfo{pages}{1149} (\bibinfo{year}{2004});
%\bibitem[{\citenamefont{Cardy}(2005)}]{Cardy-review}
\bibinfo{author}{\bibfnamefont{J.}~\bibnamefont{Cardy}}, \bibinfo{journal}{Ann.
  Phys. (N.Y.)} \textbf{\bibinfo{volume}{318}}, \bibinfo{pages}{81}
  (\bibinfo{year}{2005});
%\bibitem[{\citenamefont{Gruzberg}(2006)}]{Gruzberg-review}
\bibinfo{author}{\bibfnamefont{I.~A.} \bibnamefont{Gruzberg}},
  \bibinfo{journal}{J. Phys. A: Math. Gen.} \textbf{\bibinfo{volume}{39}},
  \bibinfo{pages}{12601} (\bibinfo{year}{2006});
%\bibitem[{\citenamefont{Bauer and Bernard}(2006)}]{BauerBernard-review}
\bibinfo{author}{\bibfnamefont{M.}~\bibnamefont{Bauer}} \bibnamefont{and}
  \bibinfo{author}{\bibfnamefont{D.}~\bibnamefont{Bernard}},
  \bibinfo{journal}{Phys. Rep.} \textbf{\bibinfo{volume}{432}},
  \bibinfo{pages}{115} (\bibinfo{year}{2006}).

\bibitem[{\citenamefont{Bernard et~al.}(2006)\citenamefont{Bernard, Boffetta,
  Celani, and Falkovich}}]{Turb1}
\bibinfo{author}{\bibfnamefont{D.}~\bibnamefont{Bernard}},
  \bibinfo{author}{\bibfnamefont{G.}~\bibnamefont{Boffetta}},
  \bibinfo{author}{\bibfnamefont{A.}~\bibnamefont{Celani}}, \bibnamefont{and}
  \bibinfo{author}{\bibfnamefont{G.}~\bibnamefont{Falkovich}},
  \bibinfo{journal}{Nature Phys.} \textbf{\bibinfo{volume}{2}},
  \bibinfo{pages}{124} (\bibinfo{year}{2006});
%\bibitem[{\citenamefont{Bernard
%  et~al.}(2007{\natexlab{a}})\citenamefont{Bernard, Boffetta, Celani, and
%  Falkovich}}]{Turb2}
%\bibinfo{author}{\bibfnamefont{D.}~\bibnamefont{Bernard}},
%  \bibinfo{author}{\bibfnamefont{G.}~\bibnamefont{Boffetta}},
%  \bibinfo{author}{\bibfnamefont{A.}~\bibnamefont{Celani}}, \bibnamefont{and}
%  \bibinfo{author}{\bibfnamefont{G.}~\bibnamefont{Falkovich}},
  \bibinfo{journal}{Phys. Rev. Lett.} \textbf{\bibinfo{volume}{98}},
  \bibinfo{pages}{024501} (\bibinfo{year}{2007}{\natexlab{a}}).

\bibitem[{\citenamefont{Amoruso et~al.}(2006)\citenamefont{Amoruso, Hartmann,
  Hastings, and Moore}}]{SG1}
\bibinfo{author}{\bibfnamefont{C.}~\bibnamefont{Amoruso}},
  \bibinfo{author}{\bibfnamefont{A.~K.} \bibnamefont{Hartmann}},
  \bibinfo{author}{\bibfnamefont{M.~B.} \bibnamefont{Hastings}},
  \bibnamefont{and} \bibinfo{author}{\bibfnamefont{M.~A.} \bibnamefont{Moore}},
  \bibinfo{journal}{Phys. Rev. Lett.} \textbf{\bibinfo{volume}{97}},
  \bibinfo{pages}{267202} (\bibinfo{year}{2006});
%\bibitem[{\citenamefont{Bernard
%  et~al.}(2007{\natexlab{b}})\citenamefont{Bernard, Doussal, and
%  Middleton}}]{SG2}
\bibinfo{author}{\bibfnamefont{D.}~\bibnamefont{Bernard}},
  \bibinfo{author}{\bibfnamefont{P.~Le} \bibnamefont{Doussal}},
  \bibnamefont{and} \bibinfo{author}{\bibfnamefont{A.~A.}
  \bibnamefont{Middleton}}, \bibinfo{journal}{Phys. Rev. B}
  \textbf{\bibinfo{volume}{76}}, \bibinfo{pages}{020403(R)}
  (\bibinfo{year}{2007}{\natexlab{b}}).

\bibitem[{\citenamefont{Saberi et~al.}(2008)\citenamefont{Saberi,
  Rajabpour, and Rouhani}}]{Growth1}
\bibinfo{author}{\bibfnamefont{A.~A.} \bibnamefont{Saberi}},
  \bibinfo{author}{\bibfnamefont{M.~A.} \bibnamefont{Rajabpour}},
  \bibnamefont{and} \bibinfo{author}{\bibfnamefont{S.}~\bibnamefont{Rouhani}},
  \bibinfo{journal}{Phys. Rev. Lett.} \textbf{\bibinfo{volume}{100}},
  \bibinfo{pages}{044504} (\bibinfo{year}{2008}{\natexlab{a}});
%\bibitem[{\citenamefont{Saberi et~al.}(2008{\natexlab{b}})\citenamefont{Saberi,
%  Niry, Fazeli, Tabar, and Rouhani}}]{Growth2}
\bibinfo{author}{\bibfnamefont{A.~A.} \bibnamefont{Saberi}},
  \bibinfo{author}{\bibfnamefont{M.~D.} \bibnamefont{Niry}},
  \bibinfo{author}{\bibfnamefont{S.~M.} \bibnamefont{Fazeli}},
  \bibinfo{author}{\bibfnamefont{M.~R.~Rahimi} \bibnamefont{Tabar}},
  \bibnamefont{and} \bibinfo{author}{\bibfnamefont{S.}~\bibnamefont{Rouhani}},
  \bibinfo{journal}{Phys. Rev. E} \textbf{\bibinfo{volume}{77}},
  \bibinfo{pages}{051607} (\bibinfo{year}{2008}{\natexlab{b}}).

\bibitem[{\citenamefont{Keating et~al.}(2006)\citenamefont{Keating, Marklof,
  and Williams}}]{Chaos1}
\bibinfo{author}{\bibfnamefont{J.~P.} \bibnamefont{Keating}},
  \bibinfo{author}{\bibfnamefont{J.}~\bibnamefont{Marklof}}, \bibnamefont{and}
  \bibinfo{author}{\bibfnamefont{I.~G.} \bibnamefont{Williams}},
  \bibinfo{journal}{Phys. Rev. Lett.} \textbf{\bibinfo{volume}{97}},
  \bibinfo{pages}{034101} (\bibinfo{year}{2006});
%\bibitem[{\citenamefont{Keating et~al.}(2008)\citenamefont{Keating, Marklof,
%  and Williams}}]{Chaos2}
%\bibinfo{author}{\bibfnamefont{J.~P.} \bibnamefont{Keating}},
%  \bibinfo{author}{\bibfnamefont{J.}~\bibnamefont{Marklof}}, \bibnamefont{and}
%  \bibinfo{author}{\bibfnamefont{I.~G.} \bibnamefont{Williams}},
  \bibinfo{journal}{New J. Phys.} \textbf{\bibinfo{volume}{10}},
  \bibinfo{pages}{083023} (\bibinfo{year}{2008}).

\bibitem[{\citenamefont{Nolin and Werner}(2009)}]{Nolin}
\bibinfo{author}{\bibfnamefont{P.}~\bibnamefont{Nolin}} \bibnamefont{and}
  \bibinfo{author}{\bibfnamefont{W.}~\bibnamefont{Werner}},
  \bibinfo{journal}{J. Amer. Math. Soc.} \textbf{\bibinfo{volume}{22}},
  \bibinfo{pages}{797} (\bibinfo{year}{2009}).

\bibitem[{\citenamefont{Bauer et~al.}(2008)\citenamefont{Bauer, Bernard, and
  Kyt{\"o}l{\"a}}}]{off-sles}
\bibinfo{author}{\bibfnamefont{M.}~\bibnamefont{Bauer}},
  \bibinfo{author}{\bibfnamefont{D.}~\bibnamefont{Bernard}}, \bibnamefont{and}
  \bibinfo{author}{\bibfnamefont{K.}~\bibnamefont{Kyt{\"o}l{\"a}}},
  \bibinfo{journal}{J. Stat. Phys.} \textbf{\bibinfo{volume}{132}}
  (\bibinfo{year}{2008});
%\bibitem[{\citenamefont{Bauer et~al.}()\citenamefont{Bauer, Bernard, and
%  Cantini}}]{off-sle-bauer}
\bibinfo{author}{\bibfnamefont{M.}~\bibnamefont{Bauer}},
  \bibinfo{author}{\bibfnamefont{D.}~\bibnamefont{Bernard}}, \bibnamefont{and}
  \bibinfo{author}{\bibfnamefont{L.}~\bibnamefont{Cantini}},
  \bibinfo{note}{arXiv:0903.1023v1 [math-ph]}.

\bibitem[{\citenamefont{Landau et~al.}(1960)\citenamefont{Landau and Lifshitz}}]{Landau}
\bibinfo{author}{\bibfnamefont{L.~D.} \bibnamefont{Landau}} \bibnamefont{and} 
  \bibinfo{author}{\bibfnamefont{E.~M.} \bibnamefont{Lifshitz}},
  \emph{\bibinfo{title}{Electrodynamics of
  Continuous Media}} (\bibinfo{publisher}{Pergamon Press},
  \bibinfo{year}{1960}).

\bibitem[{\citenamefont{Kennedy}(2008)}]{Kennedy-zipper}
\bibinfo{author}{\bibfnamefont{T.}~\bibnamefont{Kennedy}}, \bibinfo{journal}{J.
  Stat. Phys.} \textbf{\bibinfo{volume}{131}}, \bibinfo{pages}{803}
  (\bibinfo{year}{2008}).

\bibitem[{\citenamefont{Marshall and Rohde}(2007)}]{Marshall}
\bibinfo{author}{\bibfnamefont{D.~E.} \bibnamefont{Marshall}} \bibnamefont{and}
  \bibinfo{author}{\bibfnamefont{S.}~\bibnamefont{Rohde}},
  \bibinfo{journal}{SIAM J. Numer. Anal.} \textbf{\bibinfo{volume}{45}},
  \bibinfo{pages}{2577} (\bibinfo{year}{2007}).

\bibitem[12]{note}
Note that, under the employed boundary condition,  the finite system
size effect comes into the problem as the limitation that the available length of $\gamma$ is finite.

\end{thebibliography}
\end{document}